\documentclass[twocolumn,prl,showpacs]{revtex4}
\usepackage{bm}
\usepackage{epsf}

\begin{document}

\title{TUNNELING SPIN-GALVANIC EFFECT}
\author{S.A.~Tarasenko}\email{tarasenko@coherent.ioffe.ru}
\author{V.I.~Perel'}
\author{I.N.~Yassievich}
\affiliation{A.F.~Ioffe Physico-Technical Institute, RAS, 194021
St.Petersburg, Russia}

\begin{abstract}
It has been shown that tunneling of spin-polarized electrons
through a semiconductor barrier is accompanied by generation of an
electric current in the plane of the interfaces. The direction of
this interface current is determined by the spin orientation of
the electrons, in particular the current changes its direction if
the spin orientation changes the sign. Microscopic origin of such
a 'tunneling spin-galvanic' effect is the spin-orbit
coupling-induced dependence of the barrier transparency on the
spin orientation and the wavevector of electrons.
\end{abstract}


\maketitle

Spin-dependent phenomena and particularly transport of
spin-polarized carriers in semiconductor heterostructures attract
a great attention~\cite{spintronics}. One of the key problems of
spintronics is a development of efficient methods of injection and
detection of spin-polarized carriers. Among various techniques
ranging from optical orientation~\cite{oo} to spin injection from
magnetic materials (see~\cite{Kreuzer,Dorpe,Fiederling,Jiang} and
references therein), a special attention is paid to the
development of non-magnetic semiconductor injectors and detectors.
Spin-orbit interaction underlying such devices couples spin states
and space motion of conduction electrons and makes possible
effects of conversion of electric current into spin orientation
and vice versa.

Generation of electric current by spin-polarized electrons was the
subject of investigations at first in bulk materials. It was shown
that scattering of a spin-polarized electron beam is asymmetrical
due to spin-orbit interaction and therefore is accompanied by
appearance of the transversal current~\cite{DP,AY}. Such anomalous
Hall effect driven by the concentration inhomogeneity of the
optically oriented electrons was proposed in Ref.~\cite{AD} and
observed on the surface of bulk AlGaAs~\cite{Bakun}.

Recently, ability of spin-polarized carriers to drive an electric
current was demonstrated in low-dimensional semiconductor systems.
It was shown that spin relaxation of the homogeneous
spin-polarized two-dimensional electron gas yields the electric
current in systems with linear in the wavevector $\bm{k}$ spin
splitting~\cite{ILP}. This effect referred to as 'spin-galvanic'
has been recently observed in GaAs, InAs, and SiGe quantum well
structures~\cite{SGE1,SGE2}.

In this paper we demonstrate the possibility of a 'tunneling
spin-galvanic' effect. We show that tunneling of spin-polarized
electrons through the semiconductor barrier is accompanied by
generation of an electric current in the plane of the interfaces.
The direction of this interface current is determined by the spin
orientation of the electrons, in particular the current changes
its direction if the spin orientation changes the sign. The
microscopic origin of the effect under study is the spin-orbit
coupling-induced dependence of the barrier transparency on the
relative orientation of the electron spin and
wavevector~\cite{Interface,bulk}.

\begin{figure}[t]
\epsfxsize=3in \epsfysize=2.4in \centering{\epsfbox{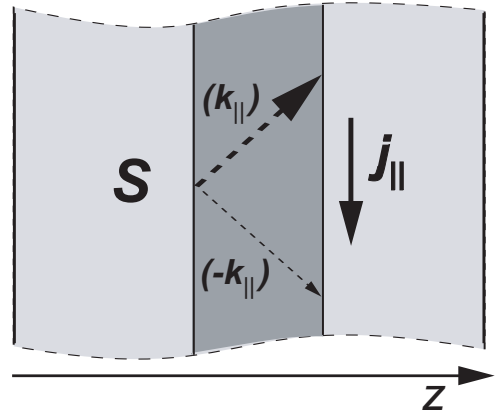}}
\caption{Origin of the tunneling spin-galvanic effect. Asymmetry
of tunneling transmission of spin-polarized carriers caused by
spin-orbit interaction results in the in-plane electric current
near the barrier.} \label{fig1}
\end{figure}

The physics of the tunneling spin-galvanic effect is sketched in
Fig.\ref{fig1}. We assume two parts of the bulk semiconductor
separated by the tunneling barrier grown along $z$ direction, and
the spin-polarized electron gas on the left side of the structure.
Spin-polarized electrons with various wavevectors tunnel through
the barrier. In the absence of spin-orbit interaction the barrier
transparency reaches maximum for the carriers propagating along
the normal to the barrier. Spin-orbit coupling changes this rule,
the optimum tunneling transmission is reached now for an oblique
incidence. The barrier transparency for the spin-polarized
carriers with the certain in-plane wavevector $\bm{k}_{\parallel}$
is large than the transparency for the particles with the opposite
in-plane wavevector, $-\bm{k}_{\parallel}$. This asymmetry results
in the in-plane flow of the transmitted electrons near the
barrier, i.e. in the interface electric current.

Generally, the barrier transparency depends on the spin
orientation of carriers if the system lacks a center of inversion.
Two microscopic mechanisms were shown to be responsible for the
effect of spin-dependent tunneling. One of them is the Rashba
spin-orbit coupling induced by the barrier
asymmetry~\cite{Interface,Zakharova,Silva,Voskoboynikov,Ting,Koga}.
The other is the $k^3$ Dresselhaus spin splitting of the electron
states in the barrier grown of a non-centrosymmetrical material
such as zinc-blende-lattice semiconductors~\cite{bulk,Botha,Hall}.
Both these mechanisms lead to the generation of the interface
current when the spin-polarized electrons tunnel through the
barrier. In the present article we consider the tunneling
spin-galvanic effect due to the Dresselhaus splitting as an
example.

The theory of the tunneling spin-galvanic effect is developed by
using the spin density matrix technique. The interface current of
spin-polarized electrons transmitted through the barrier is given
by
\begin{equation}\label{j_gen}
\bm{j}_{\parallel}=e\sum_{\bm{k}} \tau_p
\,\bm{v}_{\parallel}\,(\bm{k}) \, \mbox{Tr}
\left[\hat{g}\,(\bm{k}) \right] \:,
\end{equation}
where $e$ is the electron charge, $\tau_p$ is the momentum
relaxation time, $\bm{v}(\bm{k})$ is the velocity linked to the
electron wavevector $\bm{k}$ by the conventional expression,
$\bm{v}(\bm{k})= \hbar\bm{k}/m_1$, $m_1$ is the effective electron
mass outside the barrier, and $\hat{g}\,(\bm{k})$ is the
$2\times2$ spin matrix which describes the flux of the electrons
transmitted through the barrier. If the reverse tunneling flux
from the right to the left side of the structure is neglected then
the matrix $\hat{g}$ is determined by the electron distribution on
the left side of the barrier and the spin-dependent coefficient of
transmission, and given by
\begin{equation}\label{rhor}
\hat{g}= {\cal T} \rho_l {\cal T}^{\dag} \, v_z \Theta(v_z) \:.
\end{equation}
Here $\rho_l$ is the electron density matrix on the left side of
the structure, ${\cal T}$ is the spin matrix of the tunneling
transmission linking the incident spinor wavefunction $\psi_l$ to
the transmitted spinor wavefunction $\psi_r$, $\psi_r = {\cal T}
\psi_l$, and $\Theta$-function describes the direction of the
tunneling.

We assume the carriers on the left side of the structure to form
3D spin-oriented electron gas, and electron distributions in both
spin subband to be thermalized. Thus the density matrix has the
form
\begin{equation}
\rho_l = \frac{f_p + f_a}{2} \,\hat{I} + \frac{f_p - f_a}{2}\,
(\bm{n}_s \cdot \hat{\bm{\sigma}}) \:,
\end{equation}
where $\bm{n}_s$ is the unit vector directed along the spin
orientation, $f_p$ and $f_a$ are the distribution functions of the
electrons with the spins oriented parallel and antiparallel to
$\bm{n}_s$, respectively, and $\hat{\sigma}_{\alpha}$ are the
Pauli matrices. For the case of small degree of spin polarization,
the density matrix of 3D electron gas is simplified to
\begin{equation}\label{rhol}
\rho_l \approx f_0 \hat{I} -  \frac{d f_0}{d \varepsilon} \frac{2
p_s}{\langle 1/ \varepsilon \rangle} \,(\bm{n}_s \cdot
\hat{\bm{\sigma}}) \:,
\end{equation}
where $f_0$ is the equilibrium distribution function of
non-polarized carriers, $p_s$ is the degree of the spin
polarization, and $\langle 1/ \varepsilon \rangle$ is the average
value of the reciprocal kinetic energy of the carriers. The latter
is equal to $3/E_F$ for 3D degenerate electron gas with the Fermi
energy $E_F$, and $2/ k_B T$ and 3D non-degenerate gas at the
temperature $T$.

\begin{figure}[t]
\epsfxsize=3in \epsfysize=1.6in \centering{\epsfbox{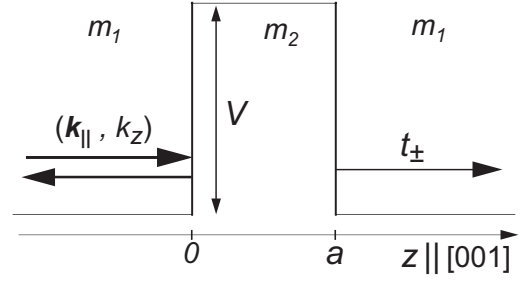}}
\caption{Tunneling through $(001)$-grown semiconductor barrier.
$V$ and $a$ are the height and the width of the barrier,
respectively.} \label{fig2}
\end{figure}

We consider the tunneling spin-galvanic effect for the symmetrical
barrier grown of a zinc-blende-lattice semiconductor along $[001]$
direction (see Fig.\ref{fig2}). In this case the barrier
transparency depends on the orientation of electron spin due to
the $k^3$ Dresselhaus spin-orbit interaction. The coefficients of
transmission for spin states "$+$" and "$-$" corresponding to the
most and the less probable tunneling have the form~\cite{bulk}
\begin{equation}\label{tpm}
t_{\pm}=t_0 \exp{ \left( \pm \, \gamma \frac{m_2
k_{\parallel}}{\hbar^2} \, aq_0  \right) } \:,
\end{equation}
where $t_0$ is the transmission coefficient when the spin-orbit
interaction is neglected, $\gamma$ is a constant of the
Dresselhaus spin-orbit coupling depending on the material, $m_2$
is the electron effective mass inside the barrier, $q_0 \approx
\sqrt{2m_2 V/\hbar^2}$ is the reciprocal length of the
wavefunction decay in the barrier, $V$ and $a$ are the height and
the width of the barrier, respectively. The orientations of the
electron spin of the states "$+$" and "$-$" depend on the
direction of the electron in-plane wavevector $\bm{k}_{\parallel}$
with respect to the crystal cubic axes. The spinors corresponding
to the spin eigen-states are given by~\cite{bulk}
\begin{equation}\label{chi_pm}
\chi_{\pm} = \frac{1}{\sqrt{2}}\left(
\begin{array}{c}
1 \\ \mp \mbox{e}^{- i \varphi}
\end{array}
\right) \:,
\end{equation}
where $\varphi$ is the polar angle of the wavevector in the $xy$
plane, being $\bm{k}_{\parallel} = (k_{\parallel} \cos \varphi \,,
\: k_{\parallel} \sin \varphi)$, and the coordinate system
$x\parallel[100]$, $y\parallel[010]$, and $z\parallel[001]$ is
assumed.

The spin matrix of the electron transmission through the barrier
is given by
\begin{equation}
{\cal T} = \sum_{s=\pm} t_{s} \,\chi_{s} \chi^{\dag}_{s} \:.
\end{equation}
For our case it has the form
\begin{equation}\label{T}
{\cal T} = \frac12 \left[
\begin{array}{cc}
t_+ + t_- & (t_- - t_+)\, e^{i\varphi} \\ (t_- - t_+) \,
e^{-i\varphi} & t_+ + t_-
\end{array} \right] \:.
\end{equation}

We assume spin corrections to be small, and the coefficient $t_0$,
for simplicity, to depend only on the $k_z$-component of the
electron wavevector. Then substituting the density
matrix~(\ref{rhol}) and the transmission matrix~(\ref{T}) into the
expressions~(\ref{j_gen},\ref{rhor}), the interface spin-dependent
current is derived to be
\begin{equation}\label{jtsge}
j_{\parallel, x} = -  j_{\parallel} \, n_{s,x} \: , \;\;
j_{\parallel, y} = j_{\parallel} \, n_{s,y} \:,
\end{equation}
\begin{displaymath}
j_{\parallel}=4 e \gamma \frac{m_2 \, a q_0}{\hbar^2}
\frac{\tau_p}{\hbar \langle 1/ \varepsilon \rangle} \dot{N} p_s
\:,
\end{displaymath}
where $\dot{N}$ is the flux of the electrons through the barrier,
$\dot{N}=\sum_{\bm{k}} \mbox{Tr}\,\hat{g} $.

The direction of the spin-dependent interface
current~(\ref{jtsge}) induced by the Dresselhaus term is
determined by the spin orientation of the electrons with respect
to the crystal axes. In particular, the current
$\bm{j}_{\parallel}$ is parallel (or antiparallel) to the spin
polarization $\bm{n}_s$, if $\bm{n}_s$ is directed along a cubic
crystal axis $[100]$ or $[010]$; and $\bm{j}_{\parallel}$ is
perpendicular to $\bm{n}_s$, if the latter is directed along the
the axis $[1 \bar{1}0]$ or $[1 10]$.

As it was mentioned above, the tunneling spin-galvanic effect can
also be induced by Rashba spin-orbit coupling in asymmetrical
barriers. In this particular case the spin-dependent interface
current flows perpendicular to the spin polarization of the
carriers.

The estimations for the tunneling spin-galvanic
current~(\ref{jtsge}) give $j_{\parallel} \sim 10^{-6} A/cm$ and
$j_{\parallel} \sim 10^{-7} A/cm$ for barriers based on GaSb and
GaAs, respectively, for the structures with the barrier
transparency $|t_0|^2 \sim 10^{-5}$ and the momentum scattering
time $\tau_p \sim 10^{-12}s$.

In conclusion, it has been demonstrated that the spin-dependent
interface current is generated if spin-polarized carriers tunnel
through the semiconductor barrier. The theory of the tunneling
spin-galvanic effect has been developed for symmetrical barriers
grown of zinc-blende-lattice compounds. The effect could be
employed for creating non-magnetic semiconductor detectors of
spin-polarized carriers.

This work was supported by the RFBR, the INTAS, and programs of
the RAS and the Russian Ministry of Industry, Science and
Technologies.

\end{document}